\documentclass{osa-article}
\journal{oe}
\articletype{Research Article}

\usepackage{lineno}

\begin{document}
	
	\title{A theoretical framework for the Hamiltonian of angular momentum optomechanical system}
	\author{Yuan Liu,\authormark{1} Dongxiao Li\authormark{1,2,*}, Yimou Liu\authormark{3}}
	\address{\authormark{1}School of Physics, Institute for Quantum Science and Engineering, Huazhong University of Science and Technology, Wuhan 430074, China}
	\address{\authormark{2}College of physics science and technology, Shenyang Normal University, Shenyang 110034, China}
	\address{\authormark{3}Center for Quantum Sciences and School of Physics, Northeast Normal University, Changchun, 130024, China}

	\email{\authormark{*}lidongxiao414@hust.edu.cn}

	\begin{abstract}
		Photon carries linear momentum and angular momentum simultaneously. Within the light-matter interaction process, exchange of linear momentum results in optical forces, whereas exchange of angular momentum leads to optical torques. Use of optical forces (light pressure or damping) have been long and wide in quantum optomechanics, however, those of optical torque and optical angular momentum are not. Here we propose a theoretical framework based on optical angular momentum flux and optical torques to derive the Hamiltonians of cavity orbital and spin angular momentum optomechanical systems, respectively. 
		Moreover, based on the method, we successfully obtain the Hamiltonian of the complex angular momentum optomechanical systems consisting of micro-cavity and several torsional oscillators, whose reflection coefficients are non-unit. Our results indicate the general applicability of our theoretical framework for the Hamiltonian of angular momentum optomechanical systems and extend the research scope of quantum optomechanics.
	\end{abstract}
	
	\section{Introduction}
	Quantum cavity optomechanics has 
	attracted lots of focus recently, ranging from applied science in
	high-sensitivity metrology \cite{Gavartin2012,PhysRevA.82.061804,Basiri-Esfahani2019} and quantum information process  \cite{PhysRevLett.102.020501,PhysRevLett.105.220501,Forsch2020,Chen2020,liu } 
	to basic science, such as the ground state cooling of mechanical oscillator \cite{Chan2011}, squeezing of both the optical \cite{PhysRevX.3.031012,Aggarwal2020} and
	the mechanical mode \cite{PhysRevA.102.053505,PhysRevLett.124.023601} and quantum entanglement at macroscopic scale \cite{Thomas2021,Kotler622,Riedinger2018}. 
	However, in almost all the optomechanical (OM) systems mentioned above, the OM interaction is induced by the linear momentum exchange interaction between light and matter. 
	
	It is well-known that in addition to exchange linear momentum with matter, light can also exchange orbital angular momentum and spin angular momentum \cite{allen2016optical} with
	matter \cite{PhysRevA.66.063402,PhysRevA.54.1593,PhysRevA.75.063409}, thus inducing orbital and spin angular momentum OM interaction, respectively. 
	This opens the door to the new possibility in the engineering of OM system.
	For example, the orbital angular momentum of Laguerre-Gaussian (LG) can be used to control the torsional motion of spiral phase plates (SPPs) \cite{PhysRevLett.99.153603}, and the spin angular momentum of light enables researchers to drive the rotational motion of levitated nano-particles \cite{PhysRevLett.117.123604,PhysRevLett.121.033602,Ahn2020} and integrated optical waveguides \cite{Fenton:18}.
	
	The quantum properties of OM systems are always determined by the OM coupling strength $g$.
	For linear momentum OM systems, $g$ can be easily derived from cavity frequency shift \cite{RevModPhys.86.1391} (i.~e.~$g\propto d\omega/dx$)  or cavity dissipation shift \cite{PhysRevLett.107.213604,PhysRevLett.110.153606} (i.~e.~$g \propto d\kappa/dx$). Nevertheless, the calculation methods of $g$ involved in angular momentum OM systems still face some difficulties. For instance, in LG beam based orbital angular momentum OM systems, it is difficult to obtain $g$ utilizing angular momentum conservation law \cite{PhysRevLett.99.153603},  
	with considering the non-unit reflection coefficient of SPP. Besides,
	for nano-particle based spin angular momentum OM systems \cite{PhysRevLett.117.123604,PhysRevLett.121.033602,Ahn2020}, only when the geometrical size of anisotropic nano-particle is smaller than the wavelength of incident light, electric dipole approximation can be utilized to
	calculation $g$.

	In this work, we propose an alternative theoretical framework based on optical angular momentum flux \cite{Barnett_2002}, aiming
	to calculate the Hamiltonian of angular momentum OM systems. The OM coupling strength of the LG beam based orbital angular momentum OM system is obtained successfully via our method, which can be further extended to more realistic and complex systems. Moreover, we also derive the Hamiltonian of spin angular momentum OM system. This Hamiltonian exhibits the cross-coupling between two orthogonal polarization modes and the torsional mode of torsional oscillator, which reflects the feature of optical angular momentum flux \cite{Barnett_2002}.
	Finally, we discuss the Hamiltonian of multi-scatters spin angular momentum OM system and 
	estimate
	the torque and angle displacement sensitivity of this system.

	This paper is organized as follows. In Sec.~\ref{general}, we derive the general Hamiltonian describing angular momentum OM system. In Sec.~\ref{orbit}, we utilize our method to obtain the SPP-based orbital angular momentum OM Hamiltonian and discuss the orbital angular momentum OM coupling strength in detail. In Sec.~\ref{spin}, we investigate the general properties of spin angular momentum OM system and demonstrate the basic application of sensing. A summary is given in Sec.~\ref{summary}.
	
	\section{Calculations of the angular momentum optomechanical Hamiltonian}
	\label{general}
	In order to understand the physical origin of 
	optomechanical interaction better, we use optical angular momentum flux to calculate the mechanical torque experienced by an optical thin membrane. According to Maxwell equations, for a monochromatic optical field, the corresponding mechanical torque $\tau_{\mathrm{mech}}$ is 
	\begin{equation}
	\mathbf{\tau}_{\mathrm{mech}}=
	-\int_{\partial V}\stackrel{\leftrightarrow}{\mathbf{M}} \cdot \mathbf{n}\mathrm{d} S,
	\label{MS}
	\end{equation}
	where the optical angular momentum flux density $\stackrel{\leftrightarrow}{\mathbf{M}}$ is defined as   $\stackrel{\leftrightarrow}{\mathbf{M}}=\mathbf{r} \times \stackrel{\leftrightarrow}{\mathbf{T}}$ with $\stackrel{\leftrightarrow}{\mathbf{T}}$ the well known Maxwell tensor \cite{novotny2012principles}.

	We consider a general model: a dielectric plate with width $D$ and infinite area, is illuminated by a laser beam with wavelength $\lambda$, which is placed in a Fabry–Pérot cavity with length $L$. In order to calculate the torque experienced by the plate, we assume the width of this dielectric plate is much smaller than the length of cavity $L$, which means $D \ll L$. This approximation guarantees that we needn't consider the electromagnetic energy trapped in the dielectric plate when we calculate the total electromagnetic energy of the cavity. We only need to consider the contribution of the $zz$ component $M_{zz}$ of the optical angular momentum flux $\stackrel{\leftrightarrow}{\mathbf{M}}$  since the area of the plate is infinity. Eq.~(\ref{MS}) shows that the mechanical torque $\tau_{z}$ is
	\begin{equation}
	\iint dxdy[ M_{zz}(x,y,-\dfrac{D}{2}) -M_{zz}(x,y,\dfrac{D}{2}) ] =\tau_{z},
	\label{1}
	\end{equation}
	which is the heart of our method to calculate the angular momentum Hamiltonian. When the 
	dielectric membrane is at position $z=0$, the corresponding OM Hamiltonian can be obtained as
	\begin{equation}
	H_{\text{angular}}^{\text{OM}}=-\hat{\tau}_{z}\hat{\theta}
	\label{OM H}.
	\end{equation}
	
	The photon Hamiltonian is $H_{\text{photon}}=\hbar\sum_{i}\omega_{i}\hat{a}_{i}^{\dagger}\hat{a}_{i}$ with the resonance frequency $\omega_{i}$ for optical mode $\hat{a}_{i}$. For the rotation motion of the mechanical oscillator with resonance frequency $\Omega$, we introduce a phonon annihilation operators $\hat{b}$ to express the mechanical Hamiltonian as $H_{\text{rot}}=\hbar\Omega \hat{b}^{\dagger}\hat{b}$. The angular displacement operator $\hat{\theta}$ can be quantized as $\hat{\theta}=\theta_{zp}(\hat{b}+\hat{b}^{\dagger})$ with $\theta_{zp}$ the zero point displacement of angular motion. Therefore, the total angular momentum Hamiltonian is
	\begin{equation}
	\begin{aligned}
 	H_{\text{angular}}&
 	=H_{\text{photon}}+H_{\text{rot}}+H_{\text{angular}}^{\text{OM}}\\
 	&=
 	\hbar\sum_{i}\omega_{i}\hat{a}_{i}^{\dagger}\hat{a}_{i}+\hbar\Omega \hat{b}^{\dagger}\hat{b} -\theta_{zp}\hat{\tau}_{z}(\hat{b}+\hat{b}^{\dagger}).
	\end{aligned}
	\label{Hangular}
	\end{equation}
	Based on Eqs.~(\ref{OM H}) and (\ref{Hangular}), one can obtain the angular momentum Hamiltonian for the various OM systems.
	In order to demonstrate the effectiveness of our method, in following sections we will discuss orbital angular momentum OM system and spin angular momentum OM system, respectively.

	\section{Orbital angular momentum optomechanical system}
	\label{orbit}
\begin{figure}
		\centering
		\includegraphics[width=\linewidth]{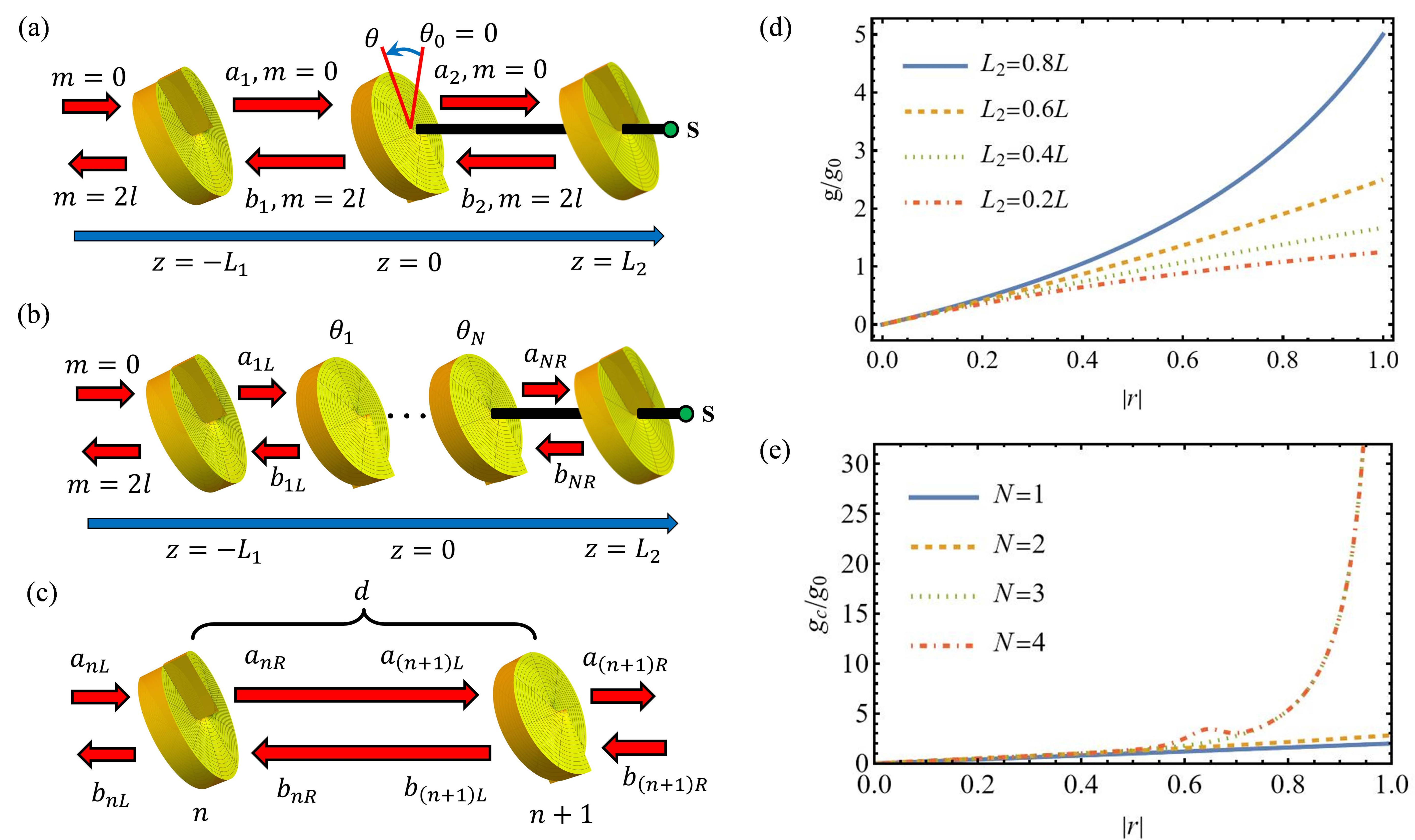}
		\caption{Orbital angular momentum OM system. (a) Single scatter system. In this system, the membrane is a SPP with spatial dependent permittivity. This plate will change the angular quantum number of input beam, thus inducing orbital angular momentum OM interaction. Here $m$ represents the topological charge of optical field. (b) and (c) are the sketches of multi-Scatters system. The distance between two adjacent SPPs is $d$. $a_{nL}$ ($a_{nR}$) and $b_{nL}$ ($b_{nR}$) are used to represent the electric field coefficients on the left (right) of the membrane $n$. (d) Relative coupling strength $g/g_{0}$ of the orbital angular momentum OM system. The relevant parameter $L$ is set as  $900\lambda$. 
			(e) Relative coupling strength $g_{c}/g_{0}$ of the multi-scatters orbital angular momentum OM system when $d=1.4\lambda$. The strength of $g$ depends on several parameters, such $d/\lambda$, $r$ and $N$. With the increasing of reflection coefficient $|r|$ and the number of scatter $N$, $g_{c}/g_{0}$ will also increase and if we choose these system parameters properly, the $g_{c}$ will much larger than $g_{0}$. In our calculation, we choose $L=900\lambda$. }
		\label{fig:Orbit}
	\end{figure}

	In this section, we will calculate the orbital angular momentum OM Hamiltonian. This system was first considered in Ref.~\cite{PhysRevLett.99.153603}, where angular momentum conservation law is used to derive the OM coupling strength and the sketch of this system is shown in Fig.~\ref{fig:Orbit}(a). Here, the yellow plate is a SPP that can change the orbital angular momentum of the input beam. The left and the right SPP constitute a resonator which is similar to Fabry–Pérot cavity. The middle spiral phase plate is supported by a torsional oscillator S. In this system, the middle SPP will change the orbital angular momentum of the input beam, thus inducing the orbital angular momentum OM interaction. Besides, $L$ is the length of this cavity, $L_{1}$ is the distance between left SPP and the middle SPP, and $L_{2}$ is the distance between right SPP and the middle SPP.

	\subsection{Single Scatters System}
	
	Firstly, we consider the Hamiltonian of  Fig.~\ref{fig:Orbit}(a). Noting that the SPP used here can cause the reflected Gaussian beam to obtain $2l$ topological charge, which is different from the usually SPP.
	In order to demonstrate the physical principle of the orbital
	angular momentum OM interaction clearly,
	we only consider the reflection losses caused by SPP and utilize a simple wave-transfer matrix \cite{PhysRevA.79.053810} 
	\begin{equation}\label{M3}
	M=\dfrac{1}{t}\begin{pmatrix}
	t^{2}-r^{2}  & r \\ 
	-r         & 1
	\end{pmatrix},
	\end{equation}
	where $r$ and $t$ are the reflection coefficient and transmission coefficient of 
	SPP, respectively, and 
	we assume that $r=|r|$, $t=i|t|$ and $|r|^{2}+|t|^{2}=1$. In our system, we assume the frequency of intra-cavity electromagnetic (E.M.) is in the range of visible spectrum and infrared spectrum. And other characteristic frequencies, such as mechanical frequency $\Omega$, mechanical dissipation rate $\Gamma$ and optical decaying rate $\kappa$, are smaller than  the order of $\mathrm{MHz}$. Thus we can utilize rotating-wave approximation \cite{PhysRevA.68.013806} to neglect the creation operator part of the intra-cavity E.M. field. Then we can obtain that:
	\begin{subequations}
		\begin{equation}
		\mathcal{E}_{1}=G[a_{1}F_{0}(\rho)e^{ikz}+b_{1}F_{2l}(\rho)e^{i2l\phi}e^{-ikz}]\hat{a}\vec{e}_{y},
		\end{equation}
		\begin{equation}
		\mathcal{E}_{2}=G[a_{2}F_{0}(\rho)e^{ikz}+b_{2}F_{2l}(\rho)e^{i2l\phi}e^{-ikz}]\hat{a}\vec{e}_{y},
		\end{equation}
		\begin{equation}
		\mathcal{B}_{1}=\dfrac{G}{c}[a_{1}F_{0}(\rho)e^{ikz}-b_{1}F_{2l}(\rho)e^{i2l\phi}e^{-ikz}]\hat{a}\vec{e}_{x},
		\end{equation}
		\begin{equation}
		\mathcal{B}_{2}=\dfrac{G}{c}[a_{2}F_{0}(\rho)e^{ikz}-b_{2}F_{2l}e^{i2l\phi}(\rho)e^{-ikz}]\hat{a}\vec{e}_{x},
		\end{equation}
	\end{subequations}
	where $F_{2l}(\rho)$  is the radial mode function of the LG beam with $\iint d\rho d\phi \rho |F_{2l}(\rho)|^{2}=1$, $2l$ represents the topological charge of LG beam, $k$ is the wave number of the intra-cavity and $a_{1},a_{2},b_{1},b_{2}$ are the relative amplitudes of the intra-cavity electric fields. $\mathcal{E}_{1}(\mathcal{B}_{1})$ represents the electric (magnetic) field at the range $-L_{1}<z<0$ and $\mathcal{E}_{2}(\mathcal{B}_{2})$ represents the electric (magnetic) field at the range $0<z<L_{2}$. Here $G$ is the normalization factor which satisfies the following equation
	\begin{equation}
	\label{N}
	G^{2}=\dfrac{2\hbar\omega}{\epsilon_{0}[L_{1}(|a_{1}|^{2}+|b_{1}|^{2})+L_{2}(|a_{2}|^{2}+|b_{2}|^{2})]}.
	\end{equation}
	Seeing Appendix~\ref{nor} for more details. 
	Besides, it is worth noting that even though the intra-cavity electric field is paraxial beam, the angular momentum flux only involves transversal E.M. components \cite{Barnett_2002}. Thereby, we do not need the exact mathematical expressions of the longitudinal E.M. fields here.

	Then we calculate the total angular momentum flux which goes through the SPP.
	On the left and right sides of the SPP, we respectively have
	\begin{subequations}
		\begin{equation}
		\iint M_{z z}(-D/2)dxdy =
		\left(\dfrac{\epsilon_{0}L}{\hbar \omega}G^2|b_{1}|^{2}\right)\hbar\dfrac{c}{L}l\hat{a}^{\dagger}\hat{a},
		\end{equation}
		\begin{equation}
		\iint M_{z z}(+D/2)dxdy =\left(\dfrac{\epsilon_{0}L}{\hbar \omega}G^2|b_{2}|^{2}\right)\hbar\dfrac{c}{L}l\hat{a}^{\dagger}\hat{a}.
		\end{equation}  
	\end{subequations}
	Here $\epsilon_{0}$ is vacuum permittivity. 
	According to Eq.~(\ref{1}), we know that the torque experienced by the middle SPP is 
	\begin{equation}
	\hat{\tau}_{z}=\iint M_{z z}(-D/2)dxdy-\iint M_{z z}(+D/2)dxdy=\left(\dfrac{\epsilon_{0}L}{\hbar \omega}G^2(|b_{1}|^{2}-|b_{2}|^{2})\right)\hbar\dfrac{c}{L}l\hat{a}^{\dagger}\hat{a}.
	\label{LG Torque}
	\end{equation}
	Substituting Eq.~(\ref{LG Torque}) into Eq.~(\ref{Hangular}), 
	we have
	\begin{equation}
	H_{\text{angular}} =\hbar \omega \hat{a}^{\dagger}\hat{a}+\hbar\Omega\hat{b}^{\dagger}\hat{b}-\hat{\tau}_{z}\hat{\theta}=\hbar \omega \hat{a}^{\dagger}\hat{a}+\hbar\Omega\hat{b}^{\dagger}\hat{b}-\hbar g \hat{a}^{\dagger}\hat{a} (\hat{b}+\hat{b}^{\dagger}),
	\label{hom}
	\end{equation}
	where 
	$g$ is
	the single photon coupling strength and 
	\begin{equation}
	g/g_{0}=\dfrac{|b_{1}|^{2}-|b_{2}|^{2}}{\dfrac{L_{1}}{L}\cdot \dfrac{(|a_{1}|^{2}+|b_{1}|^{2})}{2} +
		\dfrac{L_{2}}{L}\cdot \dfrac{(|a_{2}|^{2}+|b_{2}|^{2})}{2}}.
	\label{LGg}
	\end{equation}
	Here $g_{0}= l( c/L) \theta_{zp}$ is the orbital angular momentum coupling strength derived by the angular momentum conservation law \cite{PhysRevLett.99.153603}. 
	
	For the general situation where $L_{2}\neq 0$ and the SPP has non-unitary reflect coefficient $r$, we can also obtain the OM coupling strength $g$. Firstly,
	based on the  boundary condition $b_{2}=\exp{({i2kL_{2}})}a_{2}$ and the  wave-transfer equation
	\begin{equation}
	\begin{pmatrix}
	a_{2}\\ 
	b_{2}
	\end{pmatrix} = M\begin{pmatrix}
	a_{1}\\ 
	b_{1}
	\end{pmatrix},
	\label{scattering}
	\end{equation}
	we can obtain that \begin{subequations}
		\begin{equation}
		a_{1}=1  , 
		\end{equation}
		\begin{equation}
		b_{1}=\dfrac{  r-e^{i2kL_{2}} }{1-re^{i2kL_{2}}},   
		\end{equation}
		\begin{equation}
		a_{2}=\dfrac{ t}{1-re^{i2kL_{2}}},     
		\end{equation}
		\begin{equation}
		b_{2}=\dfrac{ te^{i2kL_{2}}}{1-re^{i2kL_{2}}}.    
		\end{equation}
	\end{subequations}
	
	Secondly, substituting these equations into Eq.~(11), we can derive the exact value of $g/g_{0}$.
	\begin{equation}
	g/g_{0}
	=\dfrac{2r(r-\cos(2kL_{2}))}{1+r^{2}-2r\cos(2kL_{2})-2r(r-\cos(2kL_{2}))(L_{2}/L)},
	\label{single g}
	\end{equation}
	which is a nonlinear function of $r$ when $\cos(2kL_{2})\neq \pm 1$ and $L_{2}/L\neq 1/2$.

	For different length $L_{2}$, the relative coupling strength $g/g_{0}$ as the function of $r$ is plotted in Fig.~\ref{fig:Orbit}(d). It is obvious that when $|r|\rightarrow1$, the middle SPP can be viewed as a perfect reflection mirror, thus $g$ will approach $L/(L-L_{2})g_{0}$. When reflection coefficient $r$ is fixed, the relative coupling strength $g/g_{0}$ will increase with the increasing of $L_{2}$. A simple explanation is that the longer $L_{2}$ is, the shorter $L_{1}$ is. Given that the photon number of the left cavity is fixed, the amplitude of the left cavity's electric field will be strengthened, hence $g/g_{0}$ will be enhanced. Thereby, the OM coupling strength $g$ will be enhanced.

	\subsection{Multi-Scatters System}
	Optomechanical coupling strength, the most vital parameter in optomechanics, can be enhanced by several methods \cite{PhysRevLett.109.223601, PhysRevA.94.053812, PhysRevA.101.033829, PhysRevLett.114.093602}, such as the introduction of the several mechanical oscillators to consider the collective mechanical motion and collective optomechanical coupling \cite{PhysRevLett.109.223601,PhysRevA.94.053812, PhysRevA.101.033829}.
	Here we extend our system to the multi-scatters OM system which is shown in Fig.~\ref{fig:Orbit}(b) and (c). Notong that in these figures, the distance between two neighbor SPP is $d$, the number of SPP is $N$ and the center SPP is in the middle of cavity.

	According to Eq.~(\ref{LG Torque}), the OM coupling strength $g_{n}$ for the $n$-th SPP is
	\begin{equation}
	g_{n}/g_{0}=\dfrac{\epsilon_{0}L}{\hbar \omega}G^2(|b_{nL}|^{2}-|b_{nR}|^{2}).
	\label{mid1}
	\end{equation}
	The coefficients $a_{nL}, b_{nL}, a_{nR}, b_{nR}$ satisfy the equations
	\begin{equation}
	\begin{pmatrix}
	a_{nR}\\ 
	b_{nR}
	\end{pmatrix} = M\begin{pmatrix}
	a_{nL}\\ 
	b_{nL}
	\end{pmatrix},
	\label{mid2}
	\end{equation}
	and
	\begin{equation}
	\begin{pmatrix}
	a_{(n+1)L}\\ 
	b_{(n+1)L}
	\end{pmatrix} =\begin{pmatrix}
	e^{ikd}	& 0 \\ 
	0	&  e^{-ikd}
	\end{pmatrix} 	
	\begin{pmatrix}
	a_{nR}\\ 
	b_{nR}
	\end{pmatrix}.
	\label{mid3}
	\end{equation}
	Following the definition of the  Ref.~\cite{PhysRevLett.109.223601}, 
	the collective coupling strength $g_{c}$ and collective motion mode $\hat{b}_{c}$ are
	\begin{equation}
	g_{c}=\sqrt{\sum_{n=1}^{N}|g_{n}|^{2}}, \quad \hat{b}_{c}=\sum_{n=1}^{N}(g_{n}^{*}/g_{c})\hat{b}_{n},
	\end{equation}
	respectively. Eq.~(\ref{single g}) has told us that when $N=1$, $g_{c}/g_{0}$ can be expressed as the form of a ratio between two polynomial functions of $r$. Furthermore, according to Eqs.~(\ref{mid1})-(\ref{mid3}), when $N>1$, we know that $g_{n}/g_{0}$ can also be expressed as the form of a ratio between two polynomial functions of $r$. Thereby, even though $g_{c}$ will increase rapidly with the increasing of $r$, it will not become exponential rise.
	
	In Fig.~\ref{fig:Orbit}(e), we exhibit the relation between the relative coupling strength $g_{c}/g_0$ and the reflection coefficient $|r|$. It is obvious that the value of $g_{c}$ can be significantly enhanced via increasing the number of the SPP. This results also certify the feasibility of our method. When $N=1$, the profile of $g/g_{0}$ will reduce to that of Fig.~1(a). When $N>1$, the profile of $g/g_{0}$ will be more complex, even will present the oscillation profile when $N=4$. The reason is that when $N>1$, photons will be reflected by SPPs for several times, and these photons will interact with each other. Consequently, $g/g_{0}$ will demonstrate a tendency of oscillation when $N=4$.
	
	\subsection{Experimental feasibility}
	Here we discuss the experimental feasibility of the orbital angular momentum optomechanical system. The most essential experimental element, spiral phase plate, has been widely researched and the fabrication of micro-optical spiral phase plate has also been experimentally demonstrated  \cite{8660645}. Therefore, it is feasible to make use of spiral phase plate to set up this optomechanical system. Besides, in our simulation, we choose the length of the cavity $L=900\lambda$ with $\lambda$ the wavelength. The relevant parameters, $\lambda=1$ $\mathrm{\mu m}$ and $L=0.9$ $\mathrm{mm}$, can also be easily achieved in experiments.

	\section{Spin angular momentum optomechanical system}
	\label{spin}
\begin{figure}
		\centering
		\includegraphics[width= 0.8\linewidth]{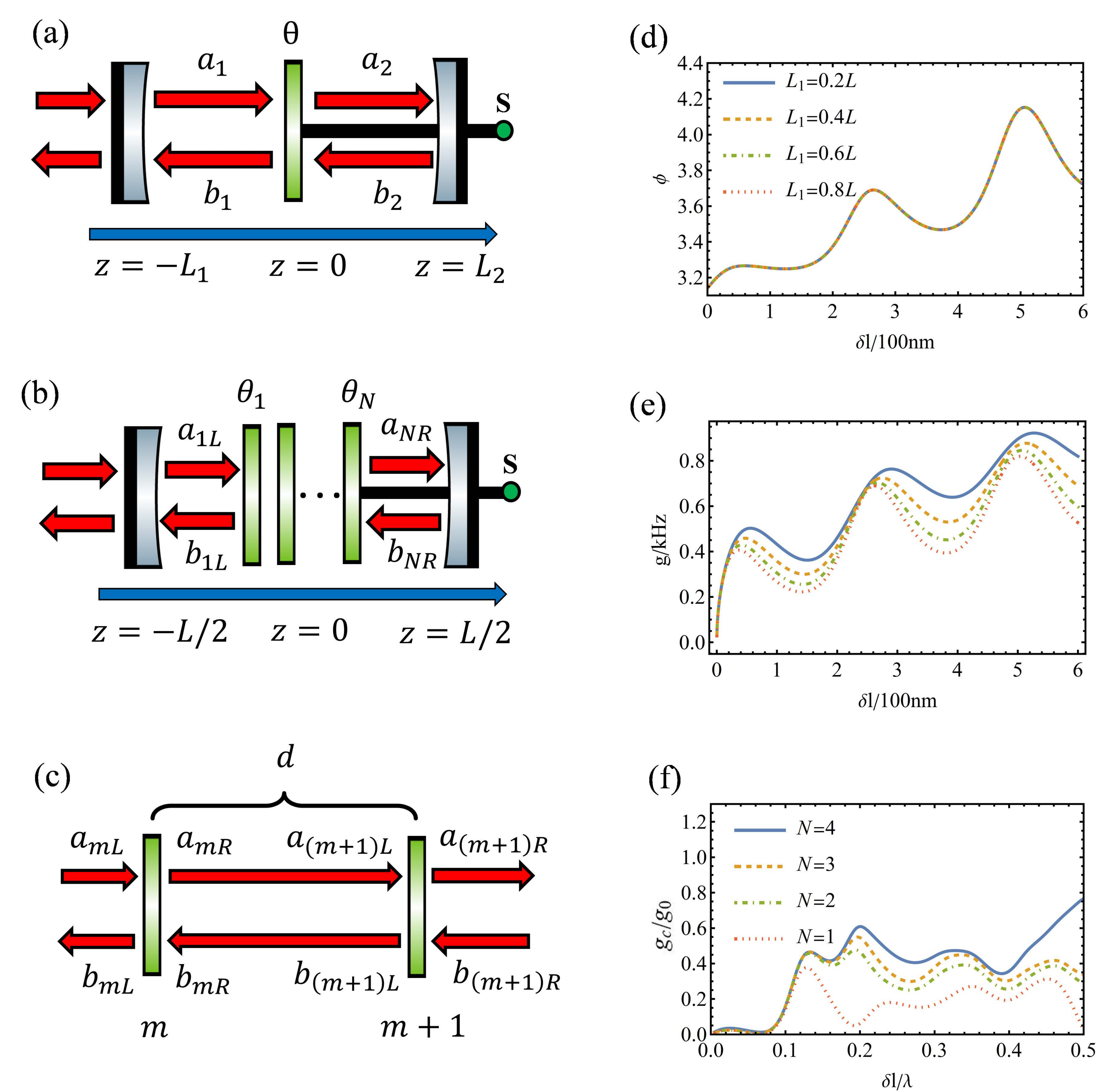}
		\caption{Spin angular momentum OM system. (a) Single scatter system. In this system, the green membrane is optical anisotropic with anisotropic permittivity tensor, thus inducing cross coupling OM coupling. (b) and (c) are the sketches of multi-Scatters system. The distance between two adjacent anisotropic permittivity is $d$. $a_{nL}$ ($a_{nR}$) and $b_{nL}$ ($b_{nR}$) are used to represent the electric field coefficients on the left (right) of the membrane $n$. The phase $\phi$ (d) and amplitude (e) of $g$ for single membrane spin angular optomechanical system. In the numerical calculation, $\lambda$ and $L$ are chosen as $780\text{nm}$ and $900\lambda$, respectively. The other parameters are listed in Table~\ref{tab2}. (f) Relative optomechanical coupling strength $g_{c} /g_{0}$ for multi-scatters spin optomechanical system when $d=1.4\lambda$. With the increasing of the number of scatter $N$, the strength of $g_{c}$ will also increase when other parameters are fixed. The related parameters are listed in Table~\ref{tab2}.}
		\label{fig:Spin}
	\end{figure}

	Even though orbital angular momentum OM system has been widely researched, it faces seriously experimental challenges, such as the high diffraction losses presented by available SPPs \cite{Shi_2016}. Thus this system has not been experimentally realized until now. Compared with optical orbital angular momentum, optical spin angular momentum is usually related with the polarization state of light and it can be manipulated more easily. In this section, we will discuss the basic physical properties and applications of spin angular momentum OM system. The sketch of this system is plotted in Fig.~\ref{fig:Spin}(a), which is similar to that of Fig.~\ref{fig:Orbit}(a) except that the middle SSP is replaced by an anisotropic optical membrane.

	\subsection{Single Scatter System}
	For plane wave, the spin angular momentum of light is related with two orthogonal circularly polarization states. In a cavity OM system, we consider two orthogonal linear polarization modes $\hat{a}_{x}$ and $\hat{a}_{y}$. The corresponding E.M. fields are
	
	\begin{subequations}
		\begin{eqnarray}
		\mathcal{E}_{x}&=&e^{im\phi}F_{m}(\rho)T_{x}(z)\hat{a}_{x},\\
		\mathcal{E}_{y}&=&e^{im\phi}F_{m}(\rho)T_{y}(z)\hat{a}_{y},\\
		\mathcal{B}_{x}&=&-\dfrac{e^{im\phi}}{i\omega}F_{m}(\rho)\dfrac{dT_{y}(z)}{dz}\hat{a}_{y},\\
		\mathcal{B}_{y}&=&\dfrac{e^{im\phi}}{i\omega}F_{m}(\rho)\dfrac{dT_{x}(z)}{dz}\hat{a}_{x}.
		\end{eqnarray}
	\end{subequations}
	Here $m$ is the topological charge of the intra-cavity E.M. field. $F_{m}(\rho)$ is the radial mode function of LG beam with $\iint d\rho d\phi|F_{m}(\rho)|^{2}=1$. $T_{x}(z)$ and $T_{y}(z)$ are the longitudinal mode functions of $\hat{a}_{x}$ and $\hat{a}_{y}$, respectively. 
	Using these equations, we can derive the operator of total optical spin angular momentum flux \cite{Barnett_2002}
	\begin{equation}\label{fluxspin3}
	\begin{aligned}
	\iint M_{z z}^{\text{spin}}dS &=\frac{\epsilon_{0} c^{2}}{2 \omega} \operatorname{Re}\left[-\mathrm{i} \iint \rho \mathrm{d} \rho \mathrm{d} \phi\left(\mathcal{E}_{x} \mathcal{B}_{x}^{*}+\mathcal{E}_{y} \mathcal{B}_{y}^{*}\right)\right]\\
	&=\frac{\epsilon_{0} c^{2}}{4 \omega^{2}} \left[
	T_{y}(z)\dfrac{dT_{x}^{*}(z)}{dz}\right]\hat{a}_{x}^{\dagger}\hat{a}_{y}-\frac{\epsilon_{0} c^{2}}{4 \omega^{2}}\left[T_{x}(z)\dfrac{dT_{y}^{*}(z)}{dz}\right]\hat{a}_{y}^{\dagger}\hat{a}_{x}
	+ h.c.
	\end{aligned}
	\end{equation}
	The optical orbital angular momentum flux \cite{Barnett_2002} is
	\begin{equation}\label{Mor}
	\begin{aligned}
	\iint M_{zz}^{\text{orbit}}dS&=m\frac{\epsilon_{0} c^{2}}{2 \omega}\operatorname{Re}\left[
	\iint\rho d\rho d\phi (-\mathcal{B}_{x}^{*}\mathcal{E}_{y}+\mathcal{E}_{x}\mathcal{B}_{y}^{*})
	\right]\\
	&=m\frac{\epsilon_{0} c^{2}}{2 \omega^{2}}
	\operatorname{Re}\left[\mathrm{i}\dfrac{dT_{y}^{*}(z)}{dz}T_{y}(z)\right]\hat{a}_{y}^{\dagger}\hat{a}_{y} 
	+m\frac{\epsilon_{0} c^{2}}{2 \omega^{2}}
	\operatorname{Re}\left[\mathrm{i}\dfrac{dT_{x}^{*}(z)}{dz}T_{x}(z)\right]\hat{a}_{x}^{\dagger}\hat{a}_{x}.
	\end{aligned}
	\end{equation}
	Noting that spin angular momentum flux $\iint M_{zz}^{\text{spin}}dS$ involves the exchange interaction between two modes $\hat{a}_{x}$ and $\hat{a}_{y}$, while orbital angular momentum flux $\iint M_{zz}^{\text{orbit}}dS$ is proportional to photon number operators $\hat{a}_{x}^{\dagger}\hat{a}_{x}$ and $\hat{a}_{y}^{\dagger}\hat{a}_{y}$.

	Firstly, we demonstrate that if the optical membrane in Fig.~\ref{fig:Spin}(a) is optical isotropic, this membrane will not experience optical torque. For simplicity, we assume that the two mode $\hat{a}_{x}$ and $\hat{a}_{y}$ are frequency degenerate with the same optical resonance frequencies $\omega$ and the same wave number $k$. Under this case, the two longitudinal mode functions $T_{x}(z)=T_{y}(z)=T(z)$ can be written as
	\begin{subequations}
		\begin{equation}
		T(z)=G(a_{1}e^{ikz}+b_{1}e^{-ikz}),\quad   -L_{1}<z<0,
		\end{equation}
		\begin{equation}
		T(z)=G(a_{2}e^{ikz}+b_{2}e^{-ikz}),\quad   0<z<+L_{2}.
		\end{equation}
	\end{subequations}
	For a single side F-P cavity, it is easy to verify that $|a_{1}|=|b_{1}|$ and $|a_{2}|=|b_{2}|$. Therefore, the spin torque $\hat{\tau}^{\text{spin}}_{z}$ is 
	\begin{equation}
	\begin{aligned}
	\hat{\tau}^{\text{spin}}_{z}&=\iint M_{zz}^{\text{spin}}(z=-D/2)dS-\iint M_{zz}^{\text{spin}}(z=D/2)dS\\
	&=\dfrac{\epsilon_{0}c^{2}k}{2\omega^{2}}\left[
	i(|b_{1}|^{2}-|a_{1}|^{2})-	i(|b_{2}|^{2}-|a_{2}|^{2})
	\right]\hat{a}^{\dagger}_{x}\hat{a}_{y}+h.c. \\& =0.
	\end{aligned}
	\end{equation}
	Substituting the exact expression of $T(z)$ into the right of Eq.~(\ref{Mor}) with $z=-D/2$, we can obtain
	\begin{equation}
	\begin{aligned}
	\operatorname{Re}\left[\mathrm{i}\dfrac{dT^{*}(z)}{dz}T(z)\right]&=k\operatorname{Re}[(a_{1}^{*}-b_{1}^{*})(a_{1}+b_{1})]=k\operatorname{Re}[a_{1}^{*}b_{1}-a_{1}b_{1}^{*}]\\
	&=2k\operatorname{Re}[\mathrm{i}\operatorname{Re}(a_{1}^{*}b_{1})]\\&=0.
	\end{aligned}
	\end{equation}
	On the right side of the membrane, we can also get the same result, so $\hat{\tau}^{\text{orbit}}_{z }=0$. Therefore, the corresponding OM Hamiltonian is
	\begin{equation}
	H_{\text{angular}}=\hbar\sum_{i=x,y}\omega_{i}\hat{a}_{i}^{\dagger}\hat{a}_{i}+\hbar\Omega \hat{b}^{\dagger}\hat{b} -\theta_{zp}\hat{\tau}_{z}(\hat{b}+\hat{b}^{\dagger})=\hbar\sum_{i=x,y}\omega_{i}\hat{a}_{i}^{\dagger}\hat{a}_{i}+\hbar\Omega \hat{b}^{\dagger}\hat{b}.
	\end{equation}
	It is obvious that at this case the spin angular momentum OM interaction strength satisfies $g=0$.

	Secondly, we try to derive the optomechanical coupling strength of Fig.~\ref{fig:Spin}(a) when the membrane is optical an isotropic.
	In the $i$ direction ($i=x,y$), we define the reflection coefficient and transmission coefficient to be $r_{i}$ and $t_{i}$, respectively. The longitudinal mode functions are
	\begin{subequations}
		\begin{equation}
		T_{\alpha}(z)=G(a_{1\alpha}e^{ikz}+b_{1\alpha}e^{-ikz}), \quad   -L_{1}<z<0,
		\end{equation}
		\begin{equation}
		T_{\alpha}(z)=G(a_{2\alpha}e^{ikz}+b_{2\alpha}e^{-ikz}), \quad   0<z<+L_{2},
		\end{equation}
	\end{subequations}
	where $\alpha=x,y$.
	Then the corresponding spin torque is
	\begin{equation}
	\hat{\tau}^{\text{spin}}_{z}
	=\iint M_{zz}^{\text{spin}}(z=-D/2)dS-\iint M_{zz}^{\text{spin}}(z=D/2)dS=g\hat{a}_{x}^{\dagger}\hat{a}_{y}+h.c., 
	\end{equation}
	and the OM coupling strength $g$ is 
	\begin{equation}
	g/g_{0}=iG_{x}G_{y}\dfrac{\epsilon_{0} L}{2\omega} [
	(b_{1x}b_{1y}^{*}-a_{1x}a_{1y}^{*})-(b_{2x}b_{2y}^{*}-a_{2x}a_{2y}^{*})],
	\label{tauspin}
	\end{equation}
	where $g_{0}=c/L\theta_{zp}$.
	Strictly speaking, $g$ is a complex number and it depends on several parameters, such as $L_{1}/L$, $r_{x}$, $r_{y}$ the width $\delta l$ of the membrane  and so on. Here we consider the membrane is a cylinder with radius $R$, width $\delta l$, mass density $\rho$ and fundamental resonance frequency $\Omega$, then the effective inertia of momentum is $I_{\text{eff}}=\pi R^{4}\rho \delta l/2$ and the zero-point angular displacement $\theta_{zp}=
	\sqrt{\hbar/(2I_{\text{eff}}\Omega)}=\sqrt{ \hbar/(\pi \Omega R^{4}\rho \delta l)}$. 
	Besides, the
	index of refraction tensor of this membrane can be written as 
	\begin{equation}
	n=\begin{pmatrix}
	n_{x}	& 0 & 0 \\ 
	0 & n_{y} & 0 \\ 
	0 & 0 & n_{x}
	\end{pmatrix}.
	\end{equation} 
	Then $r_{i}$ and $t_{i}$ ($i=x,y$) are \cite{wilson2012cavity}
	\begin{equation}
	r_{i}=\frac{\left(-n_{i}^{2}+1\right) \sin \left(k \delta l n_{i}\right)}{\left(n_{i}^{2}+1\right) \sin \left(k \delta l  n_{i}\right)+2 i n_{i} \cos \left(k \delta l n_{i}\right)},
	\end{equation}
	and
	\begin{equation}
	t_{i}=\frac{2 i n_{i}}{\left(n_{i}^{2}+1\right) \sin (k \delta l n_{i})+2 i n_{i} \cos \left(k \delta l n_{i}\right)},
	\end{equation}
	respectively. Substituting Eq.~(\ref{tauspin}) into Eq.~(\ref{Hangular}), 
	we derive the total Hamiltonian of this system 
	\begin{equation}
	H_{\text{angular}}=\hbar \omega_{x}\hat{a}_{x}^{\dagger}\hat{a}_{x} + \hbar \omega_{y}\hat{a}_{y}^{\dagger}\hat{a}_{y} + \hbar \Omega\hat{b}^{\dagger}\hat{b} + \hbar |g|(e^{i\phi}\hat{a}_{x}^{\dagger}\hat{a}_{y} + e^{-i\phi}\hat{a}_{y}^{\dagger}\hat{a}_{x} )(\hat{b}+\hat{b}^{\dagger})
	\label{hspin}
	\end{equation}
	where $\omega_{i}(i=x,y)$ is the resonance frequency of optical mode $\hat{a}_{i}$ and $\phi$ is the phase of $g$. As predicted in Eq.~(\ref{fluxspin3}), this Hamiltonian describes the exchange interaction between optical modes $\hat{a}_{x}$, $\hat{a}_{y}$ and the mechanical mode $\hat{b}$. For different width $\delta l$ and resonance frequency $\Omega$, the magnitude and phase of $g$ are plotted in Fig.~\ref{fig:Spin}(d) and Fig.~\ref{fig:Spin}(e). In Fig.~\ref{fig:Spin}(e), we observe that $g$ will oscillate with the increasing of the thickness $\delta l$. 
	Actually, this oscillation is caused by the interference effect between two optical polarization modes $\hat{a}_{x}$ and $\hat{a}_{y}$. Firstly, we note that the two optical polarization modes have  different wave number $2n_{i}\pi /\lambda  (i=x,y)$.  Secondly, because of the spin angular momentum exchange effect induced by the anisotropic optical membrane (see Eq.~(\ref{fluxspin3})), the two different polarization modes will interact with each other. Hence, the coupling strength $g$ will demonstrate the form of $\cos (2\pi (n_{x}+n_{y})\delta l/\lambda )$ or $\sin (2\pi (n_{x}+n_{y}) \delta l/\lambda)$ (also see Eq.~(\ref{tauspin})) with oscillation period $\lambda/(n_{x}+n_{y})$.

	\subsection{Multi-Scatters System}
	Here we will extend the system to multi-scatters OM system in Fig.~\ref{fig:Spin}(b) and Fig.~\ref{fig:Spin}(c).
	The coupling strength $g_{i}$ of the $n$-th membrane is
	\begin{equation}
	g_{n}/g_{0}=iG_{x}G_{y}\dfrac{\epsilon_{0} L}{2\omega}[(b_{Lnx}b_{Lny}^{*}-a_{Lnx}a_{Lny}^{*})-(b_{Rnx}b_{Rny}^{*}-a_{Rnx}a_{Rny}^{*})
	],
	\end{equation}
	where $G_{x}$ and $G_{y}$ are the corresponding normalization coefficients. We also define collective annihilation operator $\hat{b}_{\text{col}}$ and collective coupling strength $g_{\text{col}}$,
	\begin{equation}
	g_{c} =\sqrt{\sum_{n=1}^{N}|g_{n}|^{2}}, \quad \hat{b}_{c}=\sum_{n=1}^{N}(g_{n}^{*}/g_{c})\hat{b}_{n}.
	\end{equation}
	In Fig.~\ref{fig:Spin}(f), we demonstrate the relative strength $g_{c} /g_{0}$ as function of $\delta l$ with different $N$. 
	When $r_{x}$, $r_{y}$ and $d/\lambda$ are chosen some proper values, $g_{c}$ can be much larger than $g_{0}$, thereby the optomechanical coupling strength is enhanced.
	
	\subsection{Experimental feasibility}
	Compared with orbital angular momentum OM system, spin angular momentum OM system is more experimentally feasible because we only need a Fabry–Pérot cavity and an optical ansiotropic thin membrane which can be easily realized by  2-D layered material, such as $\mathrm{As_{2}S_{2}}$ membrane \cite{acsnano.9b06161} with highly optical anisotropy $n_{x}-n_{y}\sim 0.35$ and width $\delta l=129$ nm. Therefore, the parameters of our simulation is available and shown in Table~\ref{tab2}. In this table $\Gamma$ is the mechanical dissipation rate.

	\begin{table}
		\begin{center}
			\caption{Parameters of the spin angular momentum OM system}
			\begin{tabular}{ccccccc}
				\hline
				$\lambda / \mathrm{nm}$ & $\kappa/\mathrm{kHz}$ & $n_{x}$
				& $n_{y}$
				& $\Omega/ \mathrm{kHz}$ & $\Gamma/\mathrm{Hz}$ & $R/ \mathrm{\mu m}$   \\ \hline
				780 &$2\pi\times 100$&$ 1.6495$& $ 1.4822$ &$ 2\pi \times 1 $ &$3$ & 30\\
				\hline
			\end{tabular}
			\label{tab2}
		\end{center}
	\end{table}

	\subsection{Torque sensing}
	Cavity optomechanical sensor, which is one of the most important achievements in optomechanics, has emerged as a new class of ultra-precise sensors \cite{Krause2012, PhysRevLett.108.120801}, such as high-performance acceleration sensor \cite{Krause2012}, magnetic field sensor \cite{PhysRevLett.108.120801}, etc.
	Compared with angular momentum OM systems which are widely researched, the application of sensing of spin angular momentum OM systems have attracted less attention.
	With the exact spin angular momentum OM Hamiltonian derived via our method, we can 
	discuss the torque and angle sensitivity of our spin angular momentum OM system.
	
\begin{figure}
		\centering
		\includegraphics[width=1\linewidth]{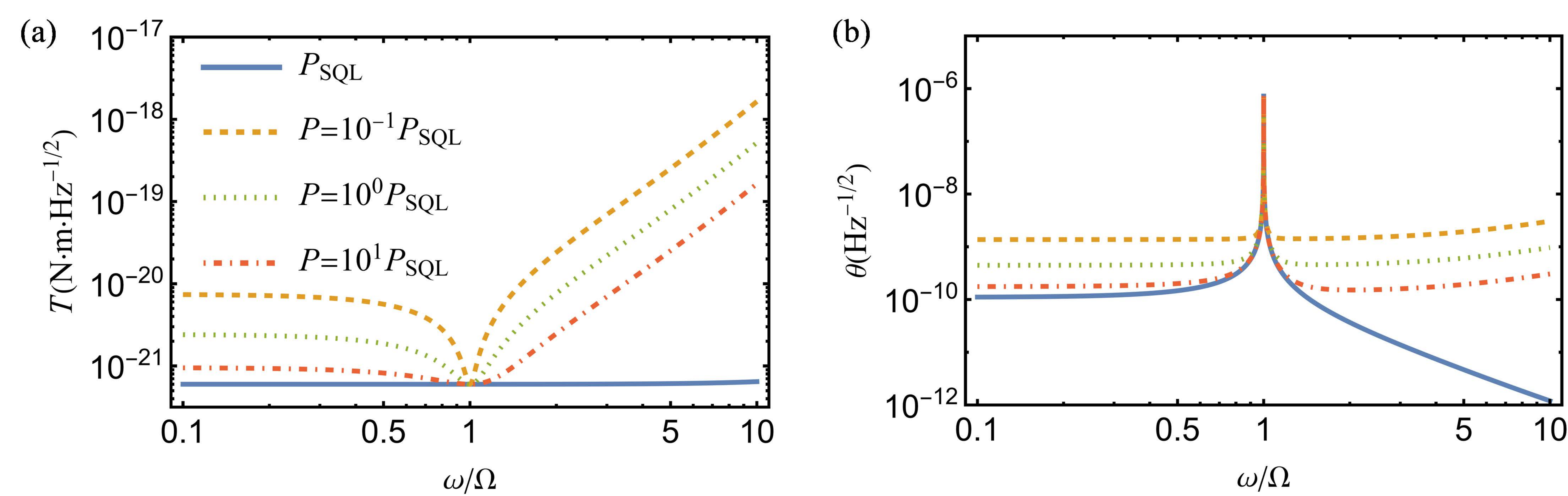}
		\caption{(a) Torque sensitivity $T$ and (b) angle sensitivity $\theta$ of our spin angular momentum system as functions of $\omega/\Omega$.}
		\label{fig:npsd}
	\end{figure}

	Firstly, we use a classical continuous laser beam with frequency $\omega_{y}$ to pump mode $\hat{a}_{y}$. The steady state average value of the intra-cavity mode $\hat{a}_{y}$ can be expressed as $\langle\hat{a}_{y} \rangle=|\alpha|e^{-i\omega_{y} t-i\phi}$. Here we adjust the phase of pumping laser to enable the phase of $\langle\hat{a}_{y} \rangle$ to be $-\phi$. If the input power of mode $\hat{a}_{y}$ is $P_{\text{in}}$, then $\alpha=\dfrac{2}{\sqrt{\kappa}}\sqrt{\dfrac{P_{\text{in}}}{\hbar \omega_{y}}}$ (see Appendix~\ref{amplitude}), where $\kappa$ is the optical dissipation rate. We also assume that the two modes $\hat{a}_{x}$ and $\hat{a}_{y}$ are degenerate with $\omega_{x}=\omega_{y}$. In the interaction picture, the OM Hamiltonian can be obtained by our method as
	\begin{equation}
	H_{\text{sense}}=\hbar\Omega \hat{b}^{\dagger}\hat{b} + \hbar |g\alpha|(\hat{a}_{x}^{\dagger} + \hat{a}_{x})(\hat{b}+\hat{b}^{\dagger}).
	\end{equation}   
	Utilizing this Hamiltonian, we can derive the power spectral density of mechanical angular displacement $S_{\theta \theta}(\omega)$ via the standard method of OM sensing \cite{bowen2015quantum}:
	\begin{equation}
	\dfrac{S_{\theta \theta}(\omega)}{2\theta_{zp}^{2}}=\frac{\omega}{Q}|\chi(\omega)|^{2}(2 \bar{n}(\omega)+1)+ \frac{1}{8\Gamma\left|C_{\mathrm{eff}}(\omega)\right|} +2 \Gamma|\chi(\omega)|^{2}\left|C_{\mathrm{eff}}(\omega)\right|,
	\end{equation}
	where $Q=\Omega/\Gamma$ is the mechanical quality factor, 
			\begin{equation}
			C_{\mathrm{eff}}(\omega) = \frac{4|g\alpha|^{2}}{(1-2 i \omega / \kappa)^{2}(\kappa \Gamma)}
			\end{equation}
			is the effective optomechanical corporation,
			\begin{equation}
			\chi(\omega)=\Omega /\left(\Omega^{2}-\omega^{2}-i \omega \Gamma\right)
			\end{equation}
			is 	the mechanical susceptibility 
			and the 
			thermal phonon number
			\begin{equation}
			\bar{n}(\omega)=\dfrac{1}{\exp(\dfrac{\hbar \omega }{k_{b}T } )-1}
			\end{equation} with $k_{b}$ the Boltzmann constant.
Besides, the standard quantum limit (SQL) of $S_{\theta \theta}(\omega)$ is 
	\begin{equation}
	\dfrac{S_{\theta \theta}^{SQL}(\omega) }{2\theta_{zp}^{2} }
	=\frac{\omega}{Q}|\chi(\omega)|^{2}(2 \bar{n}(\omega)+1)+ |\chi(\omega)|,
	\end{equation} 
	and the incident power required to reach the standard quantum limit is 
	\begin{equation}
	P_{\text{in}}^{SQL}=\frac{\hbar \omega_{y} \Gamma \kappa^{2}}{64 g^{2}}\left[1+\left(\frac{2 \Omega}{\kappa}\right)^{2}\right].
	\end{equation}
	We can also derive the torque noise power spectral density \cite{bowen2015quantum} 
	\begin{equation}
	\dfrac{S_{TT}(\omega)}{2I_{\text{eff}}\Gamma \hbar \Omega}
	=\frac{\omega}{\Omega}\left(\bar{n}(\omega)+\frac{1}{2}\right)+\frac{1}{16  \Gamma^{2}|\chi(\omega)|^{2}\left|C_{\mathrm{eff}}(\omega)\right|}+\left|C_{\text {eff }}(\omega)\right|
	\end{equation}
	and the corresponding SQL is 
	\begin{equation}
	\dfrac{S_{TT}^{SQL}(\omega) }{ 2I_{\text{eff}}\Gamma \hbar \Omega }
	=\frac{\omega}{\Omega}\left(\bar{n}(\omega)+\frac{1}{2}\right)+ \dfrac{1}{2\Gamma |\chi(\omega)|}.
	\end{equation}
	The corresponding torque sensitivity $T(\omega)=\sqrt{S_{TT}(\omega)}$ and angular position sensitivity $\theta(\omega)=\sqrt{S_{\theta \theta}(\omega)}$. In Fig.~\ref{fig:npsd}, we illustrate the torque sensitivity and angle sensitivity of our system. The used parameters are listed in Table~\ref{tab2}. Noting that when $T=3\mathrm{K}$ and $\omega\sim \Omega$, $\hbar \omega /(k_{b}T)\approx 10^{-8} \ll 1$, so  $2\bar{n}(\omega)+1\approx 2k_{b}T/(\hbar \omega)$. Besides, given that $\kappa \gg \omega $, we  have $C_{\text{eff}}(\omega)\approx C_{\text{eff}}(0) $ when $\omega\sim \Omega$. Consequently, 
			\begin{equation}
			\begin{aligned}
			\dfrac{S_{\theta \theta}(\omega)}{2\theta_{zp}^{2}}&=\frac{\omega}{Q}|\chi(\omega)|^{2}(2 \bar{n}(\omega)+1)+ \frac{1}{8\Gamma\left|C_{\mathrm{eff}}(\omega)\right|} +2 \Gamma|\chi(\omega)|^{2}\left|C_{\mathrm{eff}}(\omega)\right|\\
			&\approx \dfrac{2k_{b}T}{\hbar Q}|\chi(\omega)|^{2}+\frac{1}{8\Gamma\left|C_{\mathrm{eff}}(0)\right|}+2 \Gamma|\chi(\omega)|^{2}\left|C_{\mathrm{eff}}(0)\right|
			\end{aligned}.
			\end{equation}
			Therefore, $S_{\theta \theta}(\omega)$ is mainly decided by mechanical susceptibility $\chi(\omega)$. Considering that  $|\chi(\omega)|$ will be maximum when $\omega=\sqrt{\Omega^{2}-\Gamma^2/2}\approx \Omega$,  $\theta(\omega)=\sqrt{S_{\theta \theta}(\omega)}$ will also be maximum when $\omega\approx \Omega$. For the same reason, $T(\omega)$ will be minimum when $\omega\approx \Omega$.

	\section{Summary}
	\label{summary}
	In summary, we put forward a general theoretical framework to deal with the angular momentum OM system. For Laguerre-Gaussian beam based orbital angular momentum OM system, our method can not only re-derive the OM coupling strength obtained by
	angular momentum conservation law, but also can deal with 
	more realistic and complex orbital angular momentum OM systems, such as those with the non-unit reflection coefficient of spiral phase plate (SPP) or multiple SPPs. Moreover, for spin angular momentum OM system, the OM Hamiltonian can also be calculated by means of our method. We further obtain the torque sensitivity and the angle sensitivity of our spin angular momentum OM system. As a consequence, we believe that our method paves the road to the experimental realization of angular momentum OM system.

	\section*{Acknowledgments} 
	This work is supported by the National Natural Science Foundation of China (Grant No. 12047525), China Postdoctoral Science Foundation (Grant No. 2021M690062) and the Natural Science Foundation of Jilin Province (No. 20220101009JC).
	
	\section*{Disclosures} 
	The authors declare no conflicts of interest.

	\appendix

	\section{Normalization Factor $G$}
	\label{nor}
	Here we derive the mathematical expression of the normalization factor $G$ of Eq.~(\ref{N}). According to the expression of intra-cavity photon energy $E$, we have that
	\begin{equation}
	\begin{aligned}
	E&= \epsilon_{0}\iiint dxdydz \mathcal{E}^{*}\mathcal{E}/2\\
	&=\epsilon_{0}\iiint_{z=-L_{1}}^{z=0} dxdydz \mathcal{E}^{*}_{1}\mathcal{E}_{1}/2+\epsilon_{0}\iiint_{z=0}^{z=L_{2}} dxdydz \mathcal{E}^{*}_{2}\mathcal{E}_{2}/2\\ &= G^{2}\epsilon_{0}\left[L_{1}\left(\left|a_{1}\right|^{2}+\left|b_{1}\right|^{2}\right)+L_{2}\left(\left|a_{2}\right|^{2}+\left|b_{2}\right|^{2}\right)\right]/2.
	\end{aligned}
	\end{equation}
	Besides, the normalization regulation of quantum optics shows that photon energy $E$ should be equal with the energy of single photon $\hbar \omega$. Therefore, we have
	\begin{equation}
	G^{2}=\frac{2 \hbar \omega}{\epsilon_{0}\left[L_{1}\left(\left|a_{1}\right|^{2}+\left|b_{1}\right|^{2}\right)+L_{2}\left(\left|a_{2}\right|^{2}+\left|b_{2}\right|^{2}\right)\right]}.
	\end{equation}

	\section{Torque Sensing}
	\label{amplitude}
	In this section we will derive the expression of the steady state amplitude $\alpha$. If we use a continuous pumping field with frequency $\omega_{p}=\omega_{x}=\omega_{y}$ to pump mode $\hat{a}_{y}$, in interaction picture, the system Hamiltonian is 
	\begin{equation}
	H_{OM}=\hbar \Omega \hat{b}^{\dagger}\hat{b} + \hbar |g|(e^{i\phi}\hat{a}_{x}^{\dagger}\hat{a}_{y} + e^{-i\phi}\hat{a}_{y}^{\dagger}\hat{a}_{x} )(\hat{b}+\hat{b}^{\dagger}).
	\end{equation}
	The Langevin equation for $\hat{a}_{x}$, $\hat{a}_{y}$, $\hat{\theta}$ and $\hat{L}$ are
	\begin{subequations}
		\begin{equation}
		\dot{\hat{\theta}}=\Omega \hat{L},
		\end{equation}
		\begin{equation}
		\dot{\hat{L}}=-\Omega \hat{\theta}-\Gamma\hat{L}-\sqrt{2}|g|(e^{i\phi}\hat{a}_{x}^{\dagger}\hat{a}_{y}+h.c.)+\sqrt{2\Gamma}\hat{L}_{\text{in}},
		\end{equation}
		\begin{equation}
		\dot{\hat{a}}_{x}=-\kappa\hat{a}_{x}/2-i\sqrt{2}|g|e^{i\phi}\hat{a}_{y}\hat{\theta}+\sqrt{\kappa}\hat{a}_{x\text{in}},
		\end{equation}
		\begin{equation}
		\dot{\hat{a}}_{y}=-\kappa\hat{a}_{y}/2-i\sqrt{2}|g|e^{-i\phi}\hat{a}_{x}\hat{\theta}+\sqrt{\kappa}\hat{a}_{y\text{in}}.
		\end{equation}
	\end{subequations}
	When this system is in steady state, we have
	\begin{subequations}
		\begin{equation}
		\langle \hat{L} \rangle=0,
		\end{equation}
		\begin{equation}
		\langle \hat{a}_{x} \rangle = -i(2\sqrt{2}|g|e^{i\phi}/\kappa )\langle \hat{a}_{y} \rangle \langle \hat{\theta} \rangle
		\end{equation}
	\end{subequations}
	Then the steady state value of $\hat{\theta}$
	is
	\begin{equation}
	\langle \hat{\theta} \rangle =-\dfrac{\sqrt{2}|g|}{\Omega}(e^{i\phi}\langle \hat{a}_{x}\rangle^{*} \langle \hat{a}_{y} \rangle + c.c.)=-\dfrac{4|g|^{2}}{\Omega}(i\langle \hat{\theta} \rangle |\langle \hat{a}_{y} \rangle |^{2}+ c.c.) =0.
	\end{equation}
	As a result, the steady state value of $\hat{a}_{y}$ is 
	\begin{equation}
	\langle \hat{a}_{y} \rangle =\dfrac{2}{\sqrt{\kappa}}\langle \hat{a}_{y\text{in}} \rangle = \dfrac{2}{\sqrt{\kappa}} \sqrt{\dfrac{P_{\text{in}}}{\hbar \omega_{y}}}
	\end{equation}

	\bibliography{sample}

\end{document}